# Has the Relationship between Urban and Suburban Automobile Travel Changed across Generations? Comparing Millennials and Generation Xers in the United States

Xize Wang[a]

## Abstract

Using U.S. nationwide travel surveys for 1995, 2001, 2009 and 2017, this study compares Millennials with their previous generation (Gen Xers) in terms of their automobile travel across different neighborhood patterns. At the age of 16 to 28 years old, Millennials have lower daily personal vehicle miles traveled and car trips than Gen Xers in urban (higher-density) and suburban (lower-density) neighborhoods. Such differences remain unchanged after adjusting for the socio-economic, vehicle ownership, life cycle, year-specific and regional-specific factors. In addition, the associations between residential density and automobile travel for the 16- to 28-year-old Millennials are flatter than that for Gen Xers, controlling for the aforementioned covariates. These generational differences remain for the 24- to 36-year-old Millennials, during the period when the U.S. economy was recovering from the recession. These findings show that, in both urban and suburban neighborhoods, Millennials in the U.S. are less auto-centric than the previous generation during early life stages, regardless of economic conditions. Whether such difference persists over later life stages remains an open question and is worth continuous attention.
**Keywords:** Millennials, VMT, travel behavior, built environment, demography

a. Department of Real Estate, National University of Singapore, Singapore. Email: wangxize316@gmail.com. OCRID: 0000-0002-4861-6002

1. **Introduction**

This study investigates the relationship between built environment and automobile travel across the generations, with a focus on Millennials and their preceding generation, Generation X (Gen X). Millennials in various developed economies are less likely to travel by automobile compared with the previous generations (Kuhnimhof, Armoogum, et al., 2012). The recent nationwide travel surveys by the U.S. Federal Highway Administration, namely the National Household Travel Survey (NHTS) and Nationwide Personal Transportation Survey (NPTS), show that the 16- to 28-year-old Millennials in 2009 had 32% fewer daily car trips and 20% fewer daily personal vehicle miles traveled (VMT) than their counterparts in 1995 (see Table 1). Table 1 shows the comparison between the travel diaries from the 2009 NHTS and the 1995 NPTS. The results suggest that although reductions in driving also occurred in other age groups, the 16- to 28-year-old individuals (i.e., Millennials vs. Gen Xers) have the largest percentage difference. One important policy question is whether the existing patterns of automobile travel among Millennials will persist over time. Data from the recently-released 2017 NHTS suggest that the patterns still hold when the Millennials and Gen Xers were 24 to 36 years old (Table 2). However, whether the patterns will carry over to later life stages is still in question. To answer this question, scholars have identified several factors that influenced the travel behaviors of Millennials in the early 2010s. These factors include recession, delayed life cycles, and residence in urban neighborhoods (Blumenberg, Ralph, Smart, & Taylor, 2016; Brown, Blumenberg, Taylor, Ralph, & Voulgaris, 2016; Garikapati, Pendyala, Morris, Mokhtarian, & McDonald, 2016; McDonald, 2015; Polzin, Chu, & Godfrey, 2014; Ralph, Voulgaris, Taylor, Blumenberg, & Brown, 2016). Millennials have started to move from urban to suburban neighborhoods because of their increased income and plans for having children (Casselman, 2015). Therefore, the distances and frequencies of their automobile use are likely to increase.

Table 1 – Differences in automobility by age group (1995 – 2009)

| Age group | 1995 | 2001 | 2009 | 1995-2009 difference | difference in % | p-value |
|---|---|---|---|---|---|---|
| *Average daily personal VMT* | | | | | | |
| 16-28 | 30.3 | 29.6 | 24.4 | -5.9 | -19.51% | <0.001 |
| 29-41 | 35.0 | 35.5 | 29.9 | -5.1 | -14.55% | <0.001 |
| 42-54 | 34.2 | 34.4 | 31.0 | -3.2 | -9.34% | <0.001 |
| 55-67 | 26.8 | 29.8 | 26.9 | 0.0 | 0.11% | 0.965 |
| 68- | 17.1 | 18.9 | 16.5 | -0.7 | -3.80% | 0.258 |
| *Average daily car trips* | | | | | | |
| 16-28 | 3.8 | 3.4 | 2.6 | -1.2 | -32.01% | <0.001 |
| 29-41 | 4.2 | 4.0 | 3.3 | -0.9 | -20.95% | <0.001 |
| 42-54 | 4.1 | 4.1 | 3.5 | -0.7 | -15.78% | <0.001 |
| 55-67 | 3.7 | 3.6 | 3.2 | -0.5 | -13.66% | <0.001 |
| 68- | 2.9 | 2.9 | 2.3 | -0.6 | -19.75% | <0.001 |
| *Sample size* | | | | | | |
| 16-28 | 12,470 | 15,339 | 20,692 | | | |
| 29-41 | 18,856 | 23,566 | 28,097 | | | |
| 42-54 | 15,905 | 26,196 | 51,608 | | N/A | |
| 55-67 | 9,840 | 17,278 | 58,144 | | | |
| 68- | 8,309 | 17,329 | 53,782 | | | |

Note: Adjusted for personal weights. The p-values are from two-sample t-tests. Respondents not residing in MSAs or with daily personal VMT larger than 214 were excluded. The 16- to 28-year-old age group in 2009 was born between 1981 and 1993 and belongs to Millennials, and the 16- to 28-year-old age group in 1995 was born between 1967 and 1979 and belongs to Gen X.

Table 2 - Differences in automobility by age group (2001-2017)

| Age group | 2001 | 2009 | 2017 | 2001-2017 difference | difference in % | p-value |
|---|---|---|---|---|---|---|
| *Average personal VMT* | | | | | | |
| 24-36 | 34.0 | 28.3 | 26.2 | -7.8 | -23.0% | <0.001 |
| 37-49 | 35.6 | 31.8 | 29.4 | -6.2 | -17.3% | <0.001 |
| 50-62 | 32.4 | 28.2 | 27.1 | -5.4 | -16.6% | <0.001 |
| 63- | 20.5 | 19.3 | 20.3 | -0.2 | -1.0% | 0.662 |
| *Average daily car* | | | | | | |
| *trips* | | | | | | |
| 24-36 | 3.7 | 3.0 | 2.8 | -0.9 | -24.0% | <0.001 |
| 37-49 | 4.2 | 3.6 | 3.3 | -0.9 | -21.1% | <0.001 |
| 50-62 | 3.8 | 3.2 | 3.1 | -0.7 | -18.6% | <0.001 |
| 63- | 3.0 | 2.6 | 2.8 | -0.2 | -8.2% | <0.001 |
| *Sample size* | | | | | | |
| 24-36 | 19,168 | 20,558 | 31,448 | | | |
| 37-49 | 26,595 | 42,510 | 33,048 | | N/A | |
| 50-62 | 21,389 | 60,899 | 49,804 | | | |
| 63- | 21,099 | 73,815 | 66,957 | | | |

Note: Adjusted for personal weights. The p-values are from two-sample t-tests. Respondents not residing in MSAs or with daily personal VMT larger than 214 were excluded. The 24- to 36-year-old age group in 2017 was born between 1981-1993 and belongs to Millennials, and the 24- to 36-year-old age group in 2001 was born between 1965 and 1977 and belongs to Gen X.

However, the increasing reliance on private automobiles does not guarantee that Millennials will drive as much as the previous generations when living in similar suburban neighborhoods. Theories in demography highlight the importance of life experiences in young adulthood in shaping a generation's long-term attitudes and lifestyles (Glenn, 1980; Mannheim, 1952). Because of either the recession or their delayed life cycles, Millennials have stayed in urban neighborhoods for an extended period of time. Taking numerous short trips by public transportation or ridesharing services in young adulthood, the Millennials might develop specific activity patterns, cognitive maps and attitudes (Tal & Handy, 2010). Whenever possible, suburban Millennials may be more willing to try commuter rails, express buses, and carpooling services than previous generations. Thus, the differences in the distance and frequency of automobile travel between urban and suburban Millennials may not be as large as their counterparts in previous generations under similar socioeconomic conditions and life stages.

To test this hypothesis in the U.S. context, this study compares automobile travel among Millennials and Gen Xers in metropolitan areas across different built-environment patterns as both groups reached the same age (16-28 years old and 24-36 years old). I used the travel diary data from the four most recent nationwide travel surveys in the U.S.: the 1995 NPTS, and the 2001, 2009 and 2017 NHTSs. I first compared the changes in automobile travel patterns from 1995 to 2009 in different residential density categories for each age group. For both the urban and suburban neighborhood types, the 16- to 28-year-old Millennials had significantly lower daily personal VMT and car trips than Gen Xers of the same age range. The percentage difference in automobility for the 16- to 28-year-old individuals (i.e., Millennials vs Gen Xers) is larger than that of other age groups (i.e., other generations). I also proposed two regression models to control for the socioeconomic, vehicle ownership, life cycle, year-specific, and regional-specific factors. The results show that the marginal effects of residential density on the daily personal VMT and car trips are both significantly different between Millennials and Gen Xers. In particular, the association between residential density and daily personal VMT for the 16- to 28-year-olds in 2009 is 31% lower than those in 1995. Similarly, the association between residential density and number of daily car trips for the 16- to 28-year-olds in 2009 is 47% lower than those in 1995. I also estimated their daily personal VMT and car trips using these models holding all other covariates at their sample means, thereby adjusting for the aforementioned factors. Compared with a "covariate-adjusted Gen Xer" (in 1995), a "covariate-adjusted Millennial" (in 2009) with the same covariates had lower estimated daily personal VMT and car trips across nearly all the neighborhood density categories (except for personal VMT at the highest density level).

In addition, the 24- to 36-year-old Millennials in 2017 have higher daily personal VMT and car trips than the 16- to 28-year-old Millennials in 2001. However, the 24- to 36-year-old Millennials still have shorter and less frequent automobile trips than the Gen Xers of the same age range. The differences also apply across the residential density categories. The density-daily personal VMT and density-daily car trips associations for these 24- to 36-year-old Millennials remain significantly flatter than those of Gen Xers of the same age range. The predicted daily personal VMT and car trips for a covariate-adjusted Millennial (in 2017) are lower than a covariate-adjusted Gen Xer (in 2001) across all the residential density categories. Although these findings cannot guarantee that the travel patterns of Millennials will persist in later life stages over time, such outcomes imply that transportation planners should test various travel demand management policies on suburban Millennials.

The remainder of this paper is organized as follows. Section 2 reviews the previous literature that covers Millennial travel, theories on generations, and the relationship between the built environment and automobile travel. Section 3 introduces the dataset and the analytical methods. Section 4 reports and interprets the findings for the 16- to 28-year-old age group. Section 5 reports the analysis for the 24- to 36- year-old age group. The final section concludes this study with a brief discussion.

2. **Literature**

Since the early 2010s, researchers in transportation have observed that Millennials in developed countries, such as the U.S. (Blumenberg et al., 2016; McDonald, 2015), Canada (Marzoughi, 2011), and Germany (Kuhnimhof, Buehler, Wirtz, & Kalinowska, 2012), drive considerably less than the young adults 15–20 years ago. These Millennials were also less likely to hold a driver's license (Delbosc & Currie, 2013; Kuhnimhof, Armoogum, et al., 2012). The decrease of automobile travel among young adults (i.e., Millennials vs. Gen-Xers) is the largest compared with other age groups (McDonald, 2015). Researchers started to speculate on the contribution of this trend to the stagnant growth of automobile travel in the aforementioned period (Goodwin & Van Dender, 2013). VMT in the U.S. has recently rebounded (Federal Highway Administration, 2017) largely because of the resurgence of the economy and decrease in fuel prices (Bastian, Börjesson, & Eliasson, 2016). However, transportation planning researchers and practitioners continue to speculate whether Millennials will maintain such "multi-modalism" in the future (Circella et al., 2017; Ralph & Delbosc, 2017).

 Recent studies in the U.S. have found that the decline in automobile travel among Millennials is caused by factors such as the recent recession and the Millennials' delayed life cycle. Blumenberg et al. (2016) used data from the nationwide travel surveys conducted in 1990, 2001, and 2009 and found that unemployment was the largest contributing factor to the reduced mobility of these young adults. Garikapati et al. (2016) used data from the American Time Use Survey to argue that Millennials' delayed life cycle has caused the postponement of their adoption of the activity patterns of previous generations. Ralph et al. (2016) used a nationwide travel survey in 2009 to show that the built-environment patterns have a small, yet significant effect on Millennial driving. Polzin et al. (2014) analyzed the 2009 National Household Travel Survey and have identified the factors contributing to the decline of Millennials' automobility. These factors include the Millennials' higher educational levels, their delayed life cycles such as household formation and having children, their preference of urban/suburban areas as opposed to rural areas, and the economic recessions (Polzin et al., 2014). McDonald (2015) used nationwide travel survey data from 1995, 2001, and 2009 to quantitatively decompose the factors that contribute to the decline of automobile travel among young adults. The relative contributions of lifestyle-related demographic shifts (e.g. decreased employment), Millennial-specific factors (e.g. changing attitudes), and time-fixed effects (e.g. general dampening of the travel demand for all generations) are 10%–25%, 35%–50%, and 40%, respectively (McDonald, 2015).

Cohort theory, originally developed in demography, argues that people in different generations tend to hold unique attitudes and lifestyles (Strauss & Howe, 1997). People in a generation (or birth cohort) have experienced similar socioeconomic conditions and major historical events at similar life cycle stages (Ryder, 1965). Such collective memory, specifically in the 18–25 age range, is particularly powerful in shaping the life-long attitudes and lifestyles of people in different generations (Glenn, 1980; Mannheim, 1952). Recent studies have demonstrated the long-term impact of early adulthood experiences on people's political attitudes (Lewis-Beck, 2009), views on social policies (Alesina & Fuchs-Schündeln, 2007), and consumption patterns (Giuliano & Spilimbergo, 2009). Many Millennials have stayed in neighborhoods with good transit supply during their young adulthood for an extended period (Brown et al., 2016). They have also been exposed to a wide application of information and communication technology since an early age (Circella & Mokhtarian, 2017). Such experiences, associated with the economic hardship and the delayed life cycles that the Millennials have experienced, might be able to influence their automobile travel behavior if they reside in suburban neighborhoods at later stages of their life (Smart & Klein, 2017) by influencing their activity patterns, cognitive maps and attitudes and beliefs (Tal & Handy, 2010). Thus, the relationship between the built environment and automobile travel among Millennials may be different from that of previous generations, even with the same socioeconomic conditions and life cycle stages.

The decades-long literature on the built environment and automobile travel has shown that people residing in urban neighborhoods drive less than those in suburban neighborhoods (Ewing & Cervero, 2010). In addition, recent studies have confirmed that the built-environment patterns can affect the driving patterns after controlling for the "self-selection" effects (Cao,

Mokhtarian, & Handy, 2009). From a microeconomic perspective, urban neighborhoods have closer trip origins and destinations than their suburban counterparts. Thus, driving is less attractive in urban neighborhoods because the time and monetary costs of alternative modes are lower than those in suburban areas (Boarnet, 2011). Among the various built environment measures, density is probably the most widely-used proxy for other built environment patterns (Brownstone & Golob, 2009; Chatman, 2008; Chen, Gong, & Paaswell, 2008). In the U.S. context, residential density often correlates with land use mix and patterns of street design (Ewing & Cervero, 2010). The current literature shows that a 100% increase in the neighborhood-level residential density is associated with a 4%–12% reduction in the household-level VMT when controlling for the other built-environment factors (Ewing & Cervero, 2010). However, the reduction could be as high as 34% if no other factor is controlled for (Chatman, 2008).

The majority of the discussions on the travel behavior of young adults in the U.S. have focused on the determinants of driving, rather than the built environment – driving dynamics. Generational theory, which was developed by demographers, suggests that the relationships between urban and suburban automobile travel among young adults may differ across the generations. Accordingly, examining such potential differences can facilitate a better understanding of the current debate on the future patterns of automobile travel among Millennials, especially if they change their residential patterns.

## 3. Data and methods

*3.1. Data sources and study samples*

The dataset used in the present study comes from the public sample of the four most recent nationwide travel surveys in the U.S.: the 1995 NPTS, the 2001 NHTS, the 2009 NHTS, and the 2017 NHTS (Federal Highway Administration, 1997, 2004, 2011, 2018a). The 1995, 2001 and 2009 survey data are used to analyze the generational differences in young adulthood (16-28 years old), and the 2001, 2009 and 2017 survey data are used to test if the differences are consistent over time (24-36 years old). For the first three surveys, the sampling scheme is based on U.S. landline telephone households and is stratified by the census divisions, metropolitan area size, presence of subway/elevated rail transit systems, and two levels of telephone number density. A travel diary was mailed to each household, while personal information was acquired via telephone interviews (by interviewer in 1995 and 2001, and by computer-assisted telephone interviews in 2009). The 2017 NHTS is sampled using the addresses from postal services and stratified by the size of the metropolitan area and the existence of heavy rail transit (Federal Highway Administration, 2018a). In addition to telephone and mail, the 2017 NHTS also used the Internet to collect personal and travel diary information from some of the respondents. The response rates for the 1995, 2001, 2009, and 2017 surveys were 34.3%, 38.9%, 19.8%, and 15.6%, respectively.

Following the Pew Research Center research reports, I define Millennials as those who were born between 1981 and 1996 (Fry & Parker, 2018; Taylor, 2014). In the analyses I excluded Millennials who were less than 16 years old in 2009, because they were not yet eligible for driver's licenses at that time. Thus, the Millennials involved in this study are 16- to 28-year-old individuals in 2009 and were born between 1981 and 1993. Their counterparts in 1995 were born between 1967 and 1979 and belong to the previous generation, namely, Gen Xers. I will use the terms young adults and 16- to 28-year-old individuals interchangeably, hereafter. Hence, the first study sample consists of the 16- to 28-year-old respondents from the 1995, 2001 and 2009 surveys. In addition, I propose another 24- to 36-year-old sample from the 2001, 2009 and 2017 surveys. In this sample, the 24- to 36-year-old respondents from the 2017 survey were born between 1981 and 1993, the same as the birth years of the 16- to 28-year-old respondents in the 2009 survey. However, the 24- to 36-year-old respondents from the 2001 survey were born between 1965 and 1977; they still belong to Gen X but have slightly different birthyears than the Gen Xers in the 16- to 28-year-old sample.[1] This is because of the irregular intervals between the nationwide travel surveys. These two samples allow us to compare Millennials with Gen Xers at two different age groups: 16- to 28-year-olds and 24- to 36-year-olds.

This study proposes two variables for automobile travel: personal VMT and the number of car trips on the survey day. The data collection methods for the 2017 NHTS are slightly different from the previous surveys. I hence adjusted the 2017 travel behavior data to make it comparable with that of the 1995, 2001 and 2009 data. First, the 2017 survey makes distinctions between one-way and round trips, whereas the previous surveys only acknowledge the one-way trips. Thus, for the 2017 data, I counted the round trips twice when calculating the daily trip frequencies. Second, the 2017 survey uses an online geo-coding tool to calculate the trip distances, whereas previous surveys use self-reported trip distances. The FHWA found that the online tool systematically reports shorter trip distances than the self-reporting method does, and hence suggests inflating all the trip distances by 10% when making comparisons with earlier surveys (Federal Highway Administration, 2018b). I followed this suggestion and multiplied all the trip distances in the 2017 NHTS by 1.1. I excluded the subjects whose daily personal VMT were above 214 (the 98$^{th}$ percentiles of this variable). I also excluded the subjects who were residing outside of the metropolitan statistical areas (MSAs) to focus on the urban–suburban differences rather than the urban–rural differences. Therefore, 48,501 eligible young adults were included in the 16- to 28-year-old sample, and 71,174 eligible adults were included in the 24- to 36-year-old sample.

---

[1] Pew Research Center defines Generation X as those born between 1965 and 1980 (Taylor, 2014).

*3.2. Analytical strategy*

I used the eight residential-density categories provided by NHTS to categorize the neighborhood-level built environmental patterns. Although the residential density categories are unfortunately the only built environment variables available in the public samples of the NPTS and NHTS, these categories are usually indicators of other neighborhood design patterns in the U.S. context (Ewing & Cervero, 2010). These categories (in units of people per square mile) are below 100; 100 to 500; 500 to 1,000; 1,000 to 2,000; 2,000 to 4,000; 4,000 to 10,000; 10,000 to 25,000; and above 25,000[2]. For the 16- to 28-year-old sample, I first compared the differences in the daily personal VMT and car trips of 16- to 28-year-old individuals between 1995 and 2009 in each neighborhood category, using two-sample t-tests. Such a comparison helps to demonstrate if the reductions in automobility among young adults from 1995 to 2009 are solely caused by their different residential patterns. Thereafter, I repeated the same analysis for the other age groups (i.e., 29–41, 42–54, 55–67, and 68 and above) to show if the patterns are similar across the age groups. I used the personal weights in the comparisons to adjust for the potential biases from different sampling schemes across the surveys. To test if the generational differences in automobility across the residential patterns (with and without control variables) persist in later life stages, I repeated the above-mentioned analyses of the 16- to 28-year-old sample for the 24- to 36-year-old sample.

---

[2] Examples of the neighborhood types (unit: persons per square mile): 0 to 100, Antelope Valley, Lancaster, CA; 100 to 500, City of Industry, CA; 500 to 1,000, Friendly Hills, Mendota Heights, MN; 1,000 to 2,000, Arden Hills, MN; 2,000 to 4,000, Rancho Palos Verdes, CA; 4,000 to 10,000, Wilburton, Bellevue, WA; 10,000 to 25,000, downtown Bellevue, WA; 25,000 or above, Koreatown, Los Angeles, CA.

To account for the influence of economic factors on the direct comparisons, for each sample I also propose two regression models to compare the residential density – automobile travel associations for individuals across the generations, controlling for the socio-economic, vehicle ownership, life cycle, year-specific, and regional-specific factors. The following model descriptions use the 16- to 28-year-old sample as an illustration but also apply to the 24- to 36-year-old sample. The models follow the equation below:

$$TB_i = f(\beta_0 + \beta_1 BE_i + \mathbf{year_i} \cdot \mathbf{\beta'_2} + BE_i\, \mathbf{year_i} \cdot \mathbf{\beta'_3} + \mathbf{X_i} \cdot \mathbf{\beta'_4} + \varepsilon_i), (1)$$

where $TB_i$ is the travel behavior variable of respondent $i$ (i.e., personal VMT or number of car trips on the survey day). Notation $f(\cdot)$ indicates that neither model is an ordinary least square regression. The daily personal VMT is a left-censored variable and is estimated using the Tobit model. The daily number of car trips is a count variable and is estimated using a negative binomial model. The independent variables include an intercept term $\beta_0$, variable $BE_i$ that measures the built-environment patterns of the residence, vector $\mathbf{year_i}$ that includes the survey year (with 1995 as the reference term), and vector interacting $BE_i$ and $\mathbf{year_i}$ to separately estimate the coefficients of $BE_i$ on $TB_i$ for 1995, 2001, and 2009, as well as vector $\mathbf{X_i}$ for the control variables. Given that the public samples of these travel surveys only provide residential density categories, I constructed a continuous $BE_i$ variable using the midpoint of each category[3].

---

[3] I use 50 persons per square mile for the lowest density neighborhood category (<100 persons per square mile), and 30,000 persons per square mile for the densest neighborhood category (>25,000 persons per square mile).

The control variables in **X**$_i$ include the household income (in 2009 U.S. Dollars), driver status, number of vehicles per person in the household, age, age squared, gender, worker status, gender interacted with worker status, having children, gender interacted with having children, level of education, race of the head of household, and size of the metropolitan areas. **X**$_i$ also includes 50 state fixed-effect dummy variables to control for the inter-state variations of the 50 states and Washington D.C.[4] The time-specific variables help to control for the common trends that do not vary by location (e.g., the national trends of gasoline prices). Similarly, the regional-specific variables help to control for the common trends that do not vary by time (e.g., the baseline level of congestions). I did not use the survey weights in the regression analysis, since Winship and Radbill (1994) suggested that reporting the estimates based on unweighted data is preferred if the factors determining weights are included in the models.

Table 3 and Table 4 show the descriptive statistics of the 16- to 28-year-old and the 24- to 36-year-old samples for the regression models, respectively. These datasets are repeated cross-sectional. That is, the Federal Highway Administration (FHWA) surveyed different respondents at different survey years. Note that, for both age groups, the numbers of observations used in the regressions are smaller than those used in the two-sample t-tests. This is because the regression models have to drop individuals with missing values of control variables. I have re-run all the t-tests using these smaller "regression samples" and the signs and statistical significances of the generational differences remain unchanged.

Finally, based on the regression models, I estimated daily personal VMT and daily car trips for hypothetical Millennials and Gen Xers with the same socio-economic, vehicle ownership, life-cycle, year-specific and regional-specific covariates. Specifically, for the 16- to

---

[4] The information on the state of residence for the respondents from the 2001 NHTS comes from the geo-coded sample, since it is not available for all the respondents in the public sample.

28-year-old sample, the daily personal VMT and daily car trips were estimated for "covariate-adjusted" young adults with different residential densities in 1995 (a Gen Xer) and 2009 (a Millennial), with all the other continuous and categorical covariates set to the same values (sample means). Although using sample means produces unrealistic fractional mean values for categorical variables, it can make the predicted travel pattern variables representative of the whole sample. For the 24- to 36-year-old sample, similar estimations were made for the covariate-adjusted adults in 2001 (a Gen Xer) and 2017 (a Millennial). Such estimations can intuitively demonstrate the different density-automobility associations between the generations, with the covariates being controlled for. For instance, there are beliefs that Millennials' lower automobility is because of their delayed life stages. Such estimations can test whether Millennials still have lower levels of automobility than Gen Xers if the two had the same life-stage variables. However, such estimations are hypothetical and only useful in comparing Millennials with Gen Xers assuming that they had the same socio-demographic variables. We cannot take the estimated personal VMT or car trips as real predicted values, since the socio-demographic and life-stage characteristics will vary across different residential density categories and by generations.

Table 3 – Descriptive statistics for the 16- to 28-year-old sample for the regression models (unweighted)

| | 1995 (Gen Xers) | | 2001 (mixed) | | 2009 (Millennials) | |
|---|---|---|---|---|---|---|
| | mean | std. dev. | mean | std. dev. | mean | std. dev. |
| Personal VMT for the survey day | 30.26 | (35.63) | 29.75 | (35.03) | 25.75 | (32.12) |
| Number of car trips in the survey day | 3.81 | (3.05) | 3.43 | (2.71) | 2.73 | (2.42) |
| Population density (1,000 persons per square mile) | 6.50 | (7.99) | 5.92 | (7.56) | 4.74 | (5.88) |
| Household income (in 1,000 2009 US Dollars) | 59.79 | (39.41) | 64.83 | (40.35) | 70.06 | (38.69) |
| Driver status (1 if driver, 0 otherwise) | 0.82 | (0.38) | 0.85 | (0.35) | 0.82 | (0.38) |
| Number of vehicles per person in the household | 0.68 | (0.38) | 0.78 | (0.42) | 0.80 | (0.39) |
| Age | 22.23 | (3.97) | 21.80 | (3.98) | 20.93 | (3.98) |
| Female (1 if female, 0 otherwise) | 0.53 | (0.5) | 0.51 | (0.5) | 0.50 | (0.5) |
| Having children (=1 if having children, 0 otherwise) | 0.20 | (0.4) | 0.19 | (0.39) | 0.12 | (0.33) |
| Worker (1 if employed, 0 otherwise) | 0.72 | (0.45) | 0.76 | (0.43) | 0.55 | (0.5) |
| Education | | | | | | |
| *Less than high school\** | 0.26 | (0.44) | 0.29 | (0.45) | 0.36 | (0.48) |
| *High school/Associate degree* | 0.55 | (0.5) | 0.53 | (0.5) | 0.49 | (0.5) |
| *Bachelor's degree* | 0.16 | (0.36) | 0.15 | (0.36) | 0.12 | (0.32) |
| *Graduate degree* | 0.03 | (0.17) | 0.03 | (0.17) | 0.03 | (0.17) |
| *Not available* | 0.01 | (0.08) | 0.00 | (0.05) | 0.01 | (0.09) |
| Race of the head of household | | | | | | |
| *Non-Hispanic White* | 0.77 | (0.42) | 0.77 | (0.42) | 0.72 | (0.45) |
| *Non-Hispanic Black* | 0.09 | (0.28) | 0.06 | (0.24) | 0.07 | (0.25) |
| *Non-Hispanic Asian/Pacific Islander* | 0.03 | (0.17) | 0.04 | (0.21) | 0.04 | (0.21) |
| *Hispanic* | 0.07 | (0.25) | 0.10 | (0.3) | 0.15 | (0.36) |
| *Other races* | 0.04 | (0.19) | 0.02 | (0.14) | 0.02 | (0.14) |
| Population of metropolitan area (MSA) | | | | | | |
| *Lower than 250,000* | 0.10 | (0.3) | 0.22 | (0.42) | 0.13 | (0.34) |
| *250,000 - 499,999* | 0.08 | (0.27) | 0.22 | (0.41) | 0.10 | (0.3) |
| *500,000 - 999,999* | 0.14 | (0.35) | 0.11 | (0.32) | 0.13 | (0.34) |
| *1,000,000 - 2,999,999* | 0.23 | (0.42) | 0.15 | (0.36) | 0.26 | (0.44) |
| *3 million or higher* | 0.45 | (0.5) | 0.30 | (0.46) | 0.37 | (0.48) |
| N | 9,887 | | 14,514 | | 19,609 | |

Note: Respondents not residing in MSAs or with daily personal VMT larger than 214 were excluded.
\* The 2009 NHTS did not collect the education information for respondents younger than 18 years old, their education was imputed as *"Less than high school"*.

Table 4 – **Descriptive statistics for the 24- to 36-year-old sample for the regression models (unweighted)**

|  | 2001 (Gen Xers) | | 2009 (mixed) | | 2017 (Millennials) | |
| --- | --- | --- | --- | --- | --- | --- |
|  | mean | std. dev. | mean | std. dev. | mean | std. dev. |
| Personal VMT for the survey day | 34.06 | (37.42) | 31.00 | (34.11) | 29.22 | (34.46) |
| Number of car trips in the survey day | 3.76 | (2.78) | 3.32 | (2.68) | 3.10 | (2.52) |
| Population density (1,000 persons per square mile) | 5.82 | (7.54) | 4.86 | (6.03) | 5.62 | (6.66) |
| Household income (in 1,000 2009 US Dollars) | 69.56 | (38.08) | 68.88 | (37.11) | 73.60 | (48.97) |
| Driver status (1 if driver, 0 otherwise) | 0.94 | (0.23) | 0.94 | (0.24) | 0.93 | (0.25) |
| Number of vehicles per person in the household | 0.73 | (0.44) | 0.71 | (0.40) | 0.81 | (0.45) |
| Age | 30.66 | (3.68) | 30.82 | (3.69) | 30.45 | (3.65) |
| Female (1 if female, 0 otherwise) | 0.54 | (0.50) | 0.56 | (0.50) | 0.53 | (0.50) |
| Having children (1 if having children, 0 otherwise) | 0.55 | (0.50) | 0.56 | (0.50) | 0.37 | (0.48) |
| Worker (1 if employed, 0 otherwise) | 0.86 | (0.34) | 0.74 | (0.44) | 0.81 | (0.39) |
| Education | | | | | | |
| *Less than high school* | 0.05 | (0.23) | 0.06 | (0.23) | 0.02 | (0.15) |
| *High school/Associate degree* | 0.55 | (0.50) | 0.48 | (0.50) | 0.38 | (0.49) |
| *Bachelor's degree* | 0.28 | (0.45) | 0.30 | (0.46) | 0.35 | (0.48) |
| *Graduate degree* | 0.11 | (0.31) | 0.16 | (0.36) | 0.24 | (0.43) |
| *Not available* | 0.00 | (0.06) | 0.01 | (0.09) | 0.00 | (0.02) |
| Race of the head of household | | | | | | |
| *Non-Hispanic White* | 0.78 | (0.41) | 0.70 | (0.46) | 0.68 | (0.46) |
| *Non-Hispanic Black* | 0.05 | (0.23) | 0.07 | (0.25) | 0.07 | (0.25) |
| *Non-Hispanic Asian/Pacific Islander* | 0.05 | (0.21) | 0.05 | (0.22) | 0.08 | (0.27) |
| *Hispanic* | 0.10 | (0.30) | 0.16 | (0.37) | 0.12 | (0.33) |
| *Other races* | 0.02 | (0.14) | 0.02 | (0.14) | 0.04 | (0.21) |
| Population of metropolitan area (MSA) | | | | | | |
| *Lower than 250,000* | 0.21 | (0.41) | 0.14 | (0.34) | 0.17 | (0.38) |
| *250,000 - 499,999* | 0.22 | (0.41) | 0.10 | (0.30) | 0.11 | (0.31) |
| *500,000 - 999,999* | 0.11 | (0.31) | 0.13 | (0.34) | 0.17 | (0.38) |
| *1,000,000 - 2,999,999* | 0.15 | (0.36) | 0.26 | (0.44) | 0.19 | (0.39) |
| *3 million or higher* | 0.31 | (0.46) | 0.37 | (0.48) | 0.36 | (0.48) |
| N | 18,223 | | 19,697 | | 30,980 | |

Note: Respondents not residing in MSAs or with daily personal VMT larger than 214 were excluded.

## 4. Differences in automobility between Millennial and Gen-X young adults: the 16- to 28-year-old sample

*4.1. Generational differences in automobility by neighborhood type*

At 16–28 years old, Millennials had significantly lower daily personal VMT and car trips than their Gen-X counterparts for nearly every residential density category. The upper panel in Table 5 shows the differences in the daily personal VMTs of the 16- to 28-year-olds between 1995 (Gen-Xers) and 2009 (Millennials). For the residential density categories that range from *"below 100 persons per square mile"* to *"10,000 to 25,000 persons per square mile,"* the average daily personal VMT of Millennials is significantly lower than that of Gen Xers, with the differences ranging from 14% to 30%. The differences are significant at the 5% level, except for the *"500 to 1,000 persons per square mile"* category, which is significant at the 10% level. The difference is not significant for the *"above 25,000 persons per square mile"* category. The lower panel in Table 5 shows that the Millennial young adults took on average one car trip fewer per day than did their Gen-X counterparts in all the neighborhood categories. The differences are significant across the eight residential density categories.

**Table 5 – Differences in automobile travel (1995 – 2009) for the 16- to 28-year-old age group, by block-group level residential density**

| Residential density (persons per square mile) | 1995 (Gen Xers) | 2001 (mixed) | 2009 (Millennials) | 1995 - 2009 difference | difference in % | p-value |
|---|---|---|---|---|---|---|
| *Average daily personal VMT* | | | | | | |
| 0 to 100 | 40.5 | 40.4 | 32.7 | -7.9 | -19.5% | 0.015 |
| 100 to 500 | 42.3 | 39.7 | 33.6 | -8.7 | -20.5% | 0.001 |
| 500 to 1,000 | 33.8 | 35.9 | 29.1 | -4.6 | -13.8% | 0.082 |
| 1,000 to 2,000 | 32.1 | 32.5 | 27.4 | -4.7 | -14.5% | 0.027 |
| 2,000 to 4,000 | 32.1 | 32.2 | 24.5 | -7.6 | -23.5% | <0.001 |
| 4,000 to 10,000 | 29.2 | 28.3 | 22.4 | -6.8 | -23.2% | <0.001 |
| 10,000 to 25,000 | 22.5 | 22.2 | 15.8 | -6.7 | -29.7% | <0.001 |
| 25,000 - | 11.8 | 9.2 | 10.7 | -1.1 | -9.1% | 0.717 |
| All | 30.3 | 29.6 | 24.4 | -5.9 | -19.5% | <0.001 |
| *Average daily car trips* | | | | | | |
| 0 to 100 | 4.0 | 3.5 | 2.4 | -1.6 | -40.3% | <0.001 |
| 100 to 500 | 4.3 | 3.9 | 2.8 | -1.6 | -35.7% | <0.001 |
| 500 to 1,000 | 4.1 | 4.0 | 2.8 | -1.3 | -31.3% | <0.001 |
| 1,000 to 2,000 | 4.0 | 3.9 | 2.9 | -1.1 | -27.8% | <0.001 |
| 2,000 to 4,000 | 4.2 | 3.8 | 3.1 | -1.1 | -27.2% | <0.001 |
| 4,000 to 10,000 | 4.0 | 3.5 | 2.7 | -1.3 | -31.8% | <0.001 |
| 10,000 to 25,000 | 3.4 | 2.7 | 2.0 | -1.4 | -40.5% | <0.001 |
| 25,000 - | 1.6 | 1.2 | 0.7 | -0.9 | -58.5% | <0.001 |
| All | 3.8 | 3.4 | 2.6 | -1.2 | -32.0% | <0.001 |
| *Sample size* | | | | | | |
| 0 to 100 | 700 | 1,257 | 1,351 | | | |
| 100 to 500 | 1,617 | 2,027 | 2,880 | | | |
| 500 to 1,000 | 1,043 | 1,209 | 1,840 | | | |
| 1,000 to 2,000 | 1,397 | 1,783 | 2,846 | | N/A | |
| 2,000 to 4,000 | 2,196 | 2,719 | 4,255 | | | |
| 4,000 to 10,000 | 3,321 | 3,972 | 5,486 | | | |
| 10,000 to 25,000 | 1,348 | 1,496 | 1,560 | | | |
| 25,000 - | 848 | 876 | 474 | | | |
| All | 12,470 | 15,339 | 20,692 | | | |

Note: Adjusted for personal weights. The p-values reported are from two-sample t-tests. Respondents not residing in MSAs or with daily personal VMT larger than 214 were excluded.

Table 6 shows that the percentage differences in the average daily personal VMTs and car trips between 1995 and 2009 for the 16- to 28-year-old age group are higher than those of the other age groups. Tables 5-6 indicate that the differences in automobile travel between Millennial and Gen-X young adults occurs across various neighborhood types. A similar analysis focused on other modes show that the 16- to 28-year-old Millennials have higher frequencies of walking/biking/transit trips compared with that of Gen Xers in nearly all the neighborhood types, except for the densest. However, the magnitudes of the differences in non-motorized travel are not comparable with those for motorized trips. The results and discussion of the analysis are available in Section II of the online supplementary material.

Table 6 – Percentage differences in automobile travel (1995 - 2009), by age groups and block-group level residential density

| Residential density (persons per square mile) | 16-28 | 29-41 | 42-54 | 55-67 | 68- |
|---|---|---|---|---|---|
| *Average daily personal VMT* | | | | | |
| 0 to 100 | -19.5% | - | - | 26.2% | 28.0% |
| 100 to 500 | -20.5% | -11.2% | - | - | 21.9% |
| 500 to 1,000 | -13.8%* | -9.8%* | - | - | - |
| 1,000 to 2,000 | -14.5% | -13.9% | - | -12.6%* | - |
| 2,000 to 4,000 | -23.5% | -10.7% | -13.8% | - | -13.7% |
| 4,000 to 10,000 | -23.2% | -19.0% | -14.5% | - | - |
| 10,000 to 25,000 | -29.7% | -19.3% | - | - | - |
| 25,000 - | - | - | -53.5% | - | - |
| *Average daily car trips* | | | | | |
| 0 to 100 | -40.3% | -24.5% | -14.4% | - | - |
| 100 to 500 | -35.7% | -19.5% | -13.5% | -13.4% | - |
| 500 to 1,000 | -31.3% | -24.8% | -17.2% | -8.9%* | -19.5% |
| 1,000 to 2,000 | -27.8% | -19.6% | -8.6%* | -15.9% | - |
| 2,000 to 4,000 | -27.2% | -23.5% | -16.3% | -13.2% | -21.4% |
| 4,000 to 10,000 | -31.8% | -17.4% | -17.5% | -14.8% | -24.0% |
| 10,000 to 25,000 | -40.5% | -20.5% | -14.7% | -16.1%* | -31.0% |
| 25,000 - | -58.5% | -27.2%* | -43.3% | - | -44.0%* |

Note: Adjusted for personal weights. A negative sign indicates that the 2009 value is lower than the 1995 value, "*" indicates significance at the 10% level in a two-sample t-test, "-" indicates not significant at the 10% level in a two-sample t-test, and others indicate significance at the 5% level in a two-sample t-test. Respondents not residing in MSAs or with daily personal VMT larger than 214 were excluded. The 16- to 28-year-old age group in 2009 belong to Millennials, and the 16- to 28-year-old age group in 1995 belong to Generation X. Please refer to Table A1 in the appendix for the sample size of each group.

These differences may be influenced by several factors such as the economic conditions. The following section presents the estimates of the regression models to account for the potential influences from the socio-economic, life cycle, time-specific, and regional-specific factors. It is followed by the estimations for the expected daily personal VMT and car trips for each residential density level based on the models, thereby controlling for these factors.

*4.2. Regression models*

The two regressions in Table 7 show that daily personal VMT and number of car trips are negatively associated with residential density among young adults in the study sample. However, the association for the young adults in 2009 is significantly flatter that for the young adults in 1995 and 2001. For both models, the interaction term between residential density and the year 2009 dummy variable is significant at the 5% level. In particular, for daily personal VMT, the coefficients of density for the 1995 and 2001 young adults are both -1.043, whereas that for the 2009 group is -0.722, the summation of the coefficients of the variables *"population density"* and *"population density X 2009."* Similarly, for daily car trips, the coefficient of density for the 1995 and 2001 young adults are both -0.015, whereas that for the 2009 group is -0.008. Thus, holding other factors constant, the magnitude of the association between residential density and daily personal VMT for the young adults in 2009 was 31% lower than their counterparts in 1995. The magnitude of the association between residential density and daily car trips was 47% lower for young adults in 2009 than their counterparts in 1995.

**Table 7 – Regression models for automobile travel for the 16- to 28-year-old age group in 1995, 2001 and 2009**

| 16-28 years old | Tobit model for daily personal VMT | | Negative binomial model for daily car trips | |
|---|---|---|---|---|
| | Coef. | Std. Err. | Coef. | Std. Err. |
| Population density (1,000 people per square mile) | -1.043*** | (0.057) | -0.015*** | (0.001) |
| Density-year interactions | | | | |
| *Population density X 2001* | 0.033 | (0.073) | 0.001 | (0.002) |
| *Population density X 2009* | 0.321*** | (0.076) | 0.007*** | (0.002) |
| Survey year | | | | |
| *1995* | ref. | | ref. | |
| *2001* | -3.028*** | (0.765) | -0.172*** | (0.016) |
| *2009* | -8.903*** | (0.748) | -0.441*** | (0.016) |
| Household income (in 1,000 2009 US Dollars) | 0.047*** | (0.005) | 0.001*** | (0.000) |
| Driver status | 16.863*** | (0.608) | 0.665*** | (0.014) |
| Number of vehicles per person in the household | 10.616*** | (0.572) | 0.193*** | (0.012) |
| Age | -0.083 | (0.834) | -0.129*** | (0.018) |
| Age squared | 0.003 | (0.018) | 0.003*** | (0.000) |
| Female | 1.919*** | (0.677) | 0.087*** | (0.015) |
| Having children | 9.048*** | (0.902) | 0.141*** | (0.019) |
| Female X having children | -6.736*** | (1.063) | 0.031 | (0.022) |
| Worker | 9.750*** | (0.633) | 0.219*** | (0.014) |
| Female X worker | -1.568* | (0.809) | -0.009 | (0.018) |
| Education | | | | |
| *Less than high school* | ref. | | ref. | |
| *High school/Associate degree* | 3.615*** | (0.669) | 0.033** | (0.014) |
| *Bachelor's degree* | 6.620*** | (0.899) | 0.058*** | (0.019) |
| *Graduate degree* | 5.863*** | (1.313) | 0.052* | (0.028) |
| *Not available* | 2.986 | (2.524) | 0.110** | (0.055) |
| Race of the head of household | | | | |
| *Non-Hispanic white* | ref. | | ref. | |
| *Non-Hispanic black* | -3.913*** | (0.807) | -0.092*** | (0.017) |
| *Non-Hispanic Asian/Pacific Islander* | -1.043 | (1.019) | -0.067*** | (0.022) |
| *Hispanic* | -1.258* | (0.677) | -0.021 | (0.015) |
| *Other races* | -1.498 | (1.233) | -0.046* | (0.026) |
| Population of metropolitan area | | | | |
| *Lower than 250,000* | ref. | | ref. | |
| *250,000 - 499,999* | 1.485** | (0.734) | -0.036** | (0.015) |
| *500,000 - 999,999* | 1.532* | (0.831) | -0.039** | (0.017) |
| *1,000,000 - 2,999,999* | 0.916 | (0.716) | -0.045*** | (0.015) |
| *3 million or higher* | 0.994 | (0.712) | -0.107*** | (0.015) |
| State fixed effects | | | | |
| *Alabama* | ref. | | ref. | |
| *Alaska* | 2.322 | (6.297) | 0.181 | (0.131) |
| ... | | | | |
| *Wyoming* | -9.440 | (10.045) | 0.337* | (0.197) |
| Constant | -0.115 | (9.398) | 1.990*** | (0.200) |
| N | | 44,010 | | 44,010 |
| Pseudo R-squared | | 0.0171 | | 0.0397 |

Note: Standard errors are in parentheses, *, **, *** indicate p<0.1, p<0.05 and p<0.01, respectively. Respondents not residing in MSAs or with daily personal VMT larger than 214 were excluded. The 16- to 28-year-old in 2009 belong to Millennials, and the 16- to 28-year-old in 1995 belong to Generation X.

Similarly, the density-walking/biking/transit trip frequency association is flatter for Millennial young adults and their Gen-X counterparts. The outputs and discussion of the regression model on non-auto trips are available in Section I of the online supplementary material. In addition, the significant differences of the density–automobility relationships across generations (i.e., for daily personal VMT and daily car trips) do not vary by gender, as tested by the regression models with gender interaction terms (the outputs and discussions are available in Section II of the online supplementary material).

Table 7 also shows that many of the control variables are significantly associated with automobile travel. Being a driver and living in a household with more vehicles per person positively correlate with the distance and frequency of automobile travel among young adults. Age negatively associates with daily car trips. The magnitude of this association decreases as age increases. The marginal effects of gender, worker status, and having children on automobile travel are complicated because of their multiple interaction terms. In particular, female young adults with no work or children have higher daily personal VMT and more daily car trips than their male counterparts. For both male and female young adults, being employed and having children are positively associated with personal VMT. A high education level is positively associated with daily personal VMT and the number of car trips per day. In addition, the race of the household head correlates with automobile travel among young adults. Compared with young adults with a non-Hispanic white household head, those with a non-Hispanic African-American household head travel by car less frequently and for shorter distances; and those with a non-Hispanic Asian or Pacific Islander household head tend to have lower frequencies of car trips. Meanwhile, young adults residing in large metropolitan areas tend to have fewer car trips than those in smaller metropolitan areas. The personal VMT and metropolitan sizes of young adults are likely to follow an inverse-U shaped relationship. Lastly, the negative coefficients of the two dummy variables on the year fixed effects show the general dampening of automobile travel among young adults across all the aforementioned characteristics. The time- and state-fixed effects adjust for the time- and state-specific variables, such as gas prices and policies on driver's licenses.

*4.3. Estimated automobile travel patterns adjusted by covariates*

To intuitively demonstrate the different density–automobility associations across generations, this section estimates the daily personal VMT and car trips for a covariate-adjusted "average" young adult with different residential densities in 1995 and 2009 (see Figure 1). "Covariate-adjusted" indicates that all other control variables shown in Table 7 were all set as their mean values[5]. Figure 1 shows that, holding the socio-economic, vehicle ownership, life cycle, year-specific, and regional-specific factors constant, a covariate-adjusted young adult in a less dense neighborhood has higher personal VMT and number of car trips per day than her/his counterpart in a denser neighborhood. In addition, the differences between the urban and suburban young adults in both measures of automobility for young adults in 2009 (Millennials) are smaller than their counterparts in 1995 (Gen-Xers). The estimated daily personal VMT for the 2009 covariate-adjusted young adult is lower than that of her/his 1995 counterpart across most residential density categories. The only exception is the highest level of neighborhood density (see Figure 1[a]). The estimated numbers of car trips per day for the "covariate-adjusted" young adult in 2009 are fewer than their counterparts in 1995 across all the levels of residential density (Figure 1[b]). Thus, the findings of Section 4.1 that 16- to 28-year-old Millennials travel in automobiles for shorter distances and less frequently than their Gen X counterparts across the density categories (see Table 5) remain after adjusting for socio-economic, vehicle ownership, life cycle, year-specific, and regional-specific covariates.

---

[5] The mean values of the 50 state-fixed effect variables are not shown in Table 2 because of space constraints. These data are available upon request.

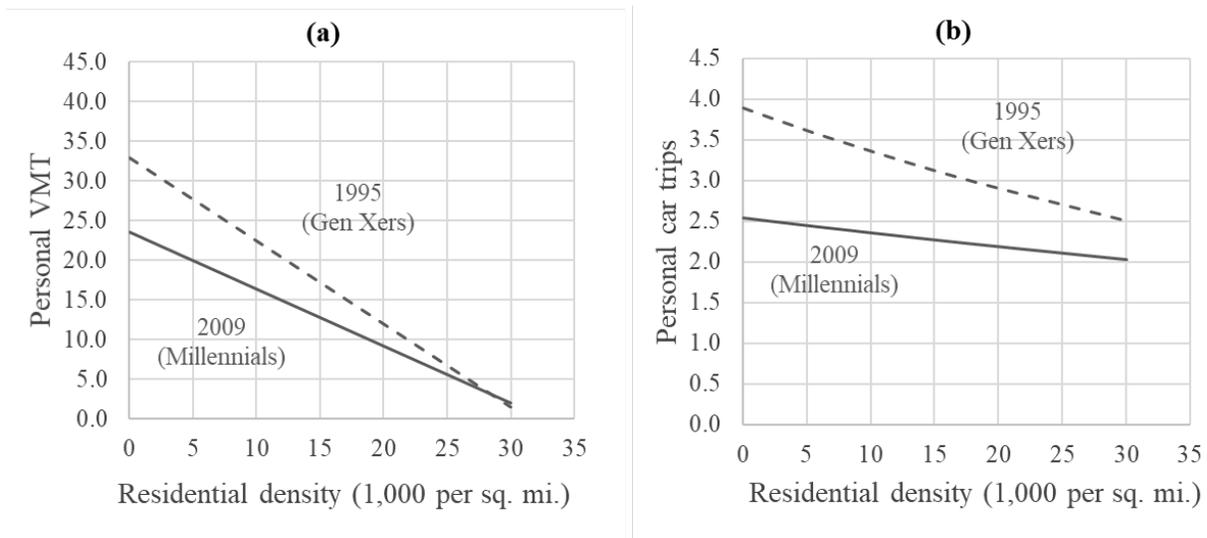

**Figure 1 – Estimated daily personal VMT and car trips for a covariate-adjusted young adult in 1995 and 2009**
(Note: refer to Section 3 for examples of each neighborhood type, all other covariates are set to their sample means from the regression models in Table 7)

In summary, the analyses show that at the same young adult age (16 to 28 years old), Millennials (1) traveled less in automobiles (in distance and frequency) than Gen Xers in all the neighborhood types but the densest and (2) have smaller differences in automobility between urban and suburban neighborhoods than Gen Xers. These findings hold with or without controlling for the socio-economic, vehicle ownership, life cycle, year-specific, and regional-specific factors.

5. **Consistent generational differences over time: the 24- to 36-year-old sample**

This section uses the three most recent NHTSs (i.e., 2001, 2009, and 2017) to show that the lower levels of automobility of the 16-to 28-year-old Millennials relative to their counterparts in Generation X still hold at the 24- to 36-year-old age range. The respondents at 24- to 36-years old in the 2017 NHTS, born between 1981 and 1993, are drawn from the same population as the 16- to 28-year-old Millennials in the 2009 NHTS. The 24- to 36-year-old Millennials in the 2017 survey travel for longer distances and more frequently in automobiles than the 16- to 28-year-old Millennials in 2009. For Millennials born between 1981 and 1993, the average daily personal VMT for the 2009 and 2017 samples are 24.4 and 26.2, respectively (a 7.4% difference, $p < 0.001$ for a two-sample t-test adjusted for survey weights). Similarly, the average daily car trips for the 2009 and 2017 samples are 2.6 and 2.8, respectively (an 8.6% difference, $p < 0.001$). However, compared with the Gen Xers of the same age (i.e., 24- to 36-years old) in 2001, these Millennials still travel for shorter distances and less frequently in automobiles (Table 8). In particular, the average daily personal VMT of the 24- to 36-year-old Millennials (26.2) is 23% lower than that of Gen Xers (34.0). Moreover, the average daily car trips (2.8) of Millennials is 24% lower than that of Gen Xers (3.7). Both differences are significant at the 1% level.

The differences in the average daily personal VMT and car trips between Millennials and Gen Xers at 16 to 28 years old across the residential density categories (see Table 5) still hold at the age range of 24-36. Table 8 shows that 24- to 36-year-old Millennials have significantly lower daily personal VMT and car trips than their Gen-X counterparts for all density categories, except for the densest. The difference in daily personal VMT at the *"below 100 persons per square mile"* category is significant at the 10% level, whereas all other differences are significant at the 5% level. Analyses on other modes for the same 24- to 36-year-old sample found significant differences in daily walking/biking/transit trips across the urban and suburban neighborhoods. However, the differences are at a smaller magnitude than that those for car trips (please refer to Section I of the supplementary material for details). In addition, Table 9 indicates that the percentage differences in automobile travel between 24- to 36-year-old Millennials and Gen Xers are larger than those of other age groups. These findings are similar to those for the 16- to 28-year-old sample shown in Table 6.

Table 8 – Differences in automobile travel (2001 – 2017) for the 24- to 36-year-old age group, by block-group level residential density

| Residential density (persons per square mile) | 2001 (Gen Xers) | 2009 (mixed) | 2017 (Millennials) | 2001 - 2017 change | change in % | p-value |
|---|---|---|---|---|---|---|
| *Average daily personal VMT* | | | | | | |
| 0 to 100 | 48.3 | 34.8 | 42.0 | -6.3 | -13.0% | 0.072 |
| 100 to 500 | 43.6 | 38.6 | 36.5 | -7.1 | -16.3% | 0.001 |
| 500 to 1,000 | 40.1 | 38.3 | 34.1 | -6.0 | -15.0% | 0.015 |
| 1,000 to 2,000 | 38.5 | 31.8 | 31.2 | -7.3 | -19.0% | <0.001 |
| 2,000 to 4,000 | 36.2 | 28.6 | 27.2 | -9.0 | -24.8% | <0.001 |
| 4,000 to 10,000 | 32.6 | 26.5 | 23.5 | -9.2 | -28.1% | <0.001 |
| 10,000 to 25,000 | 25.7 | 18.9 | 20.2 | -5.5 | -21.3% | <0.001 |
| 25,000 - | 14.0 | 13.8 | 11.6 | -2.4 | -17.1% | 0.186 |
| All | 34.0 | 28.3 | 26.2 | -7.8 | -23.0% | <0.001 |
| *Average daily car trips* | | | | | | |
| 0 to 100 | 3.9 | 2.8 | 3.3 | -0.6 | -15.6% | 0.003 |
| 100 to 500 | 4.2 | 3.3 | 3.2 | -1.0 | -23.4% | <0.001 |
| 500 to 1,000 | 4.2 | 3.1 | 3.3 | -0.9 | -22.0% | <0.001 |
| 1,000 to 2,000 | 4.2 | 3.4 | 3.1 | -1.2 | -27.2% | <0.001 |
| 2,000 to 4,000 | 4.1 | 3.4 | 3.1 | -1.0 | -24.4% | <0.001 |
| 4,000 to 10,000 | 3.8 | 3.3 | 2.9 | -0.9 | -24.5% | <0.001 |
| 10,000 to 25,000 | 3.2 | 2.4 | 2.4 | -0.8 | -23.8% | <0.001 |
| 25,000 - | 1.5 | 1.0 | 1.3 | -0.2 | -12.2% | 0.240 |
| All | 3.7 | 3.0 | 2.8 | -0.9 | -24.0% | <0.001 |
| *Sample size* | | | | | | |
| 0 to 100 | 1,455 | 1,156 | 1,685 | | | |
| 100 to 500 | 2,573 | 2,801 | 3,486 | | | |
| 500 to 1,000 | 1,616 | 1,880 | 2,659 | | | |
| 1,000 to 2,000 | 2,406 | 2,924 | 4,059 | | | |
| 2,000 to 4,000 | 3,343 | 4,228 | 6,340 | | N/A | |
| 4,000 to 10,000 | 4,960 | 5,438 | 9,052 | | | |
| 10,000 to 25,000 | 1,715 | 1,615 | 3,066 | | | |
| 25,000 - | 1,100 | 516 | 1,101 | | | |
| All | 19,168 | 20,558 | 31,448 | | | |

Note: Adjusted for personal weights. The p-values reported are from two-sample t-tests. Respondents not residing in MSAs or with daily personal VMT larger than 214 were excluded.

**Table 9 – Percentage differences in automobile travel (2001 - 2017), by age groups and block-group level residential density**

| Residential density (persons per square mile) | 24-36 | 37-49 | 50-62 | 63- |
|---|---|---|---|---|
| *Average daily personal VMT* | | | | |
| 0 to 100 | -13.0%* | -12.9% | -12.6% | 13.0%* |
| 100 to 500 | -16.3% | -11.2% | -11.5% | - |
| 500 to 1,000 | -15.0% | -17.7% | -21.4% | 21.4% |
| 1,000 to 2,000 | -19.0% | -18.5% | -21.5% | -14.7% |
| 2,000 to 4,000 | -24.8% | -17.5% | -13.2% | -10.1% |
| 4,000 to 10,000 | -28.1% | -15.6% | -14.6% | -7.4%* |
| 10,000 to 25,000 | -21.3% | -13.9% | -16.0%* | - |
| 25,000 - | - | -33.7% | - | - |
| *Average daily car trips* | | | | |
| 0 to 100 | -15.6% | -17.6% | -19.7% | -10.5%* |
| 100 to 500 | -23.4% | -18.7% | -14.4% | - |
| 500 to 1,000 | -22.0% | -17.1% | -19.4% | - |
| 1,000 to 2,000 | -27.2% | -24.4% | -19.1% | -16.3% |
| 2,000 to 4,000 | -24.4% | -23.9% | -18.4% | -11.0% |
| 4,000 to 10,000 | -24.5% | -20.0% | -19.1% | -7.8% |
| 10,000 to 25,000 | -23.8% | -17.2% | -11.6%* | - |
| 25,000 - | - | -21.3% | -34.9% | - |

Note: Adjusted for personal weights. A negative sign indicates that the 2017 value is lower than the 2001 value, "*" indicates significance at 10% level in a two-sample t-test, "-" indicates not significant at the 10% level in a two-sample t-test, others indicate significant at the 5% level in a two-sample t-test. Respondents not residing in MSAs or with daily personal VMT larger than 214 were excluded. The 24- to 36-year-old age group in 2017 belongs to Millennials, and the 24- to 36-year-old age group in 2001 belongs to Generation X. Please refer to Table A2 in the appendix for the sample size of each group.

The regression models in Table 10 show that the density–automobility association for the 24- to 36-year-old Millennials is flatter than that of the Gen-Xers, with the socio-economic, vehicle ownership, life cycle, year-specific, and regional-specific factors adjusted for. For daily personal VMT, the coefficient for the 2001 Gen-Xers is -1.072, whereas that for the 2017 Millennials is −0.772 (a 28% difference). For daily car trips, the coefficient is −0.013 for the 2001 respondents (Generation X) and −0.009 for the 2017 respondents (Millennials, a 31% difference). This finding is consistent with that in Table 7 for the 16-28-year-old sample. However, it is worth noting that the generational differences in the 24- to 36-year-old sample is smaller than those in the young adult sample (see Table 7). In 2017, the US economy has recovered from the recent recession, which was at its deepest when the 2009 NHTS was conducted. These findings using the 2017 data supports the arguments of Polzin et al. (2014) and Smart and Klein (2017) that some characteristics in travel may be consistent over time regardless of economic conditions.

Table 10 – Regression models for automobile travel for the 24- to 36-year-old age group in 2001, 2009 and 2017

| 24-36 years old | Tobit model for daily personal VMT | | Negative binomial model for daily car trips | |
| --- | --- | --- | --- | --- |
|  | Coef. | Std. Err. | Coef. | Std. Err. |
| Population density (1,000 people per square mile) | -1.072*** | (0.045) | -0.013*** | (0.001) |
| Density-year interactions |  |  |  |  |
| *Population density X 2009* | 0.128* | (0.067) | 0.003** | (0.001) |
| *Population density X 2017* | 0.215*** | (0.058) | 0.004*** | (0.001) |
| Survey year |  |  |  |  |
| *2001* | ref. |  | ref. |  |
| *2009* | -4.739*** | (0.607) | -0.165*** | (0.012) |
| *2017* | -7.798*** | (0.544) | -0.217*** | (0.010) |
| Household income (in 1,000 2009 US Dollars) | 0.032*** | (0.004) | <0.001 | (<0.001) |
| Driver status | 26.316*** | (0.790) | 0.908*** | (0.018) |
| Number of vehicles per person in the household | 10.354*** | (0.442) | 0.131*** | (0.009) |
| Age | -0.790 | (0.764) | -0.018 | (0.015) |
| Age squared | 0.012 | (0.013) | <0.001 | (<0.001) |
| Female | 3.780*** | (0.891) | 0.129*** | (0.018) |
| Having children | 9.164*** | (0.529) | 0.162*** | (0.010) |
| Female X having children | -4.466*** | (0.633) | 0.064*** | (0.012) |
| Worker | 14.835*** | (0.777) | 0.202*** | (0.016) |
| Female X worker | -3.320*** | (0.911) | -0.050*** | (0.019) |
| Education | | | | |
| *Less than high school* | ref. | | ref. | |
| *High school/Associate degree* | 2.868*** | (0.868) | 0.100*** | (0.018) |
| *Bachelor's degree* | 5.119*** | (0.911) | 0.158*** | (0.018) |
| *Graduate degree* | 3.378*** | (0.956) | 0.149*** | (0.019) |
| *Not available* | -1.164 | (2.914) | -0.064 | (0.061) |
| Race of the head of household | | | | |
| *Non-Hispanic white* | ref. | | ref. | |
| *Non-Hispanic black* | 0.003 | (0.673) | -0.006 | (0.013) |
| *Non-Hispanic Asian/Pacific Islander* | -2.918*** | (0.681) | -0.109*** | (0.014) |
| *Hispanic* | -0.202 | (0.524) | -0.005 | (0.010) |
| *Other races* | 0.132 | (0.907) | -0.015 | (0.018) |
| Population of metropolitan area | | | | |
| *Lower than 250,000* | ref. | | ref. | |
| *250,000 - 499,999* | -0.052 | (0.585) | -0.036*** | (0.011) |
| *500,000 - 999,999* | 1.694*** | (0.625) | -0.016 | (0.012) |
| *1,000,000 - 2,999,999* | 2.310*** | (0.564) | -0.017 | (0.011) |
| *3 million or higher* | 3.620*** | (0.557) | -0.081*** | (0.011) |
| State fixed effects | | | | |
| *Alabama* | ref. | | ref. | |
| *Alaska* | 1.264 | (4.775) | -0.003 | (0.093) |
| *...* | | | | |
| *Wyoming* | -16.865** | (6.555) | -0.176 | (0.129) |
| Constant | -5.194 | (11.801) | 0.203 | (0.231) |
| N | 68,900 | | 68,900 | |
| Pseudo R-squared | 0.0133 | | 0.0279 | |

Note: Standard errors in parentheses, *, **, *** indicate p<0.1, p<0.05 and p<0.01, respectively. Respondents not residing in MSAs or with personal VMT larger than 214 were excluded. The 24- to 36-year-old in 2017 belongs to Millennials, and the 24- to 36-year-old in 2001 belongs to Generation X.

Regarding biking, walking and transit trips, the density-trip frequency association is also flatter for the Millennial young adults and their Gen-X counterparts. The outputs and the discussion of the regression model are available in Section I of the online supplementary material. In addition, the significant differences in the density-daily car trips relationship between the generations do not vary by gender for the 24- to 36-year-old sample, as tested by the regression models with gender interaction terms. This pattern is consistent with that of the 16- to 28-year-old sample. However, the density-daily personal VMT associations of the female Millennials are flatter than that of the male Millennials. The outputs and detailed discussions of the regression models are available in Section II of the online supplementary material.

The estimated personal VMT and car trips adjusted by covariates for the 24- to 36-year-old sample in 2001 and 2017 show a pattern that is consistent with that for the 16- to 28-year-old sample (Figure 2). Specifically, holding the socio-economic, vehicle ownership, life cycle, year-specific, and regional-specific factors constant, a "covariate-adjusted" individual in a less dense neighborhood has higher estimated personal VMT and number of car trips per day than her/his counterpart in a denser neighborhood. In addition, the differences between the urban and suburban residents in both measures of automobility for those in 2017 (Millennials) are smaller than their counterparts in 2001 (Gen Xers). The estimated daily personal VMT and personal car trips for the 2017 covariate-adjusted adults are lower than that of their 2001 counterparts across most of the residential density levels. Thus, the aforementioned findings that 24- to 36-year-old Millennials travel in automobiles for shorter distances and less frequently than their Gen-X counterparts across all of the density categories (as shown in Table 8) remain after adjusting for the covariates.

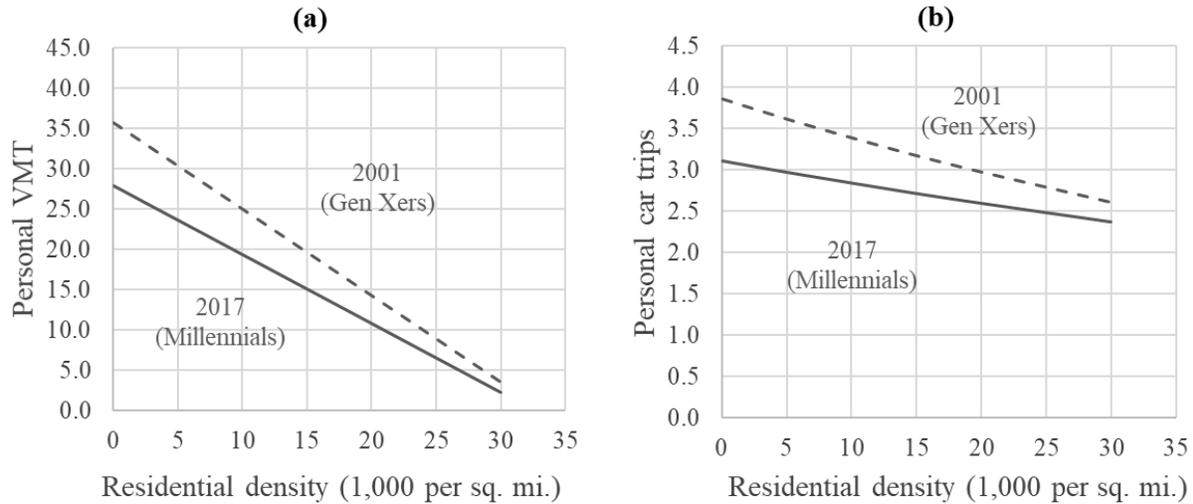

**Figure 2 – Estimated daily personal VMT and car trips for a covariate-adjusted individual in 2001 and 2017**
(Note: refer to Section 3 for examples of each neighborhood type, all other covariates are set to their sample means from the regression models in Table 10)

## 6. Conclusions

Using nationwide travel diaries from the U.S. for 1995, 2001, 2009 and 2017, this study finds that young adults aged 16 to 28 years old in 2009 traveled for shorter distances and less frequently via automobiles than their counterparts in 1995 when they were residing in urban and suburban neighborhoods with similar residential densities. The regression models show that such difference still holds even when the socioeconomic, vehicle ownership, life cycle, year-specific, and regional-specific factors are controlled for. The regression models also show that the density−daily personal VMT and density−daily car trips associations are flatter among the 2009 young adults (Millennials) than their 1995 counterparts (Gen Xers). These generational differences in automobility for the 16- to 28-year-old sample still hold for the 24- to 36-year-old sample. The 24- to 36-year-old Millennials travel for longer distances and more frequently in automobiles in 2017 than the 16- to 28-year-old Millennials in 2009. However, in all but the densest neighborhood categories, they consistently have lower daily personal VMT and car trips than Gen Xers at both the 16- to 28-year-old and 24- to 36-year-old ranges, regardless of the economic conditions.

Admittedly, the persistence of the relationship between urban and suburban travel among Millennials in later life stages remains unclear. On one hand, life stage is an important factor shaping people's travel behavior (Scheiner & Holz-Rau, 2013). It is possible that Millennials will catch up with earlier generations when catching up their delayed life stages (e.g. forming families and raising children), as argued by Newbold and Scott (2017) for Canada's Millennials. One the other hand, staying in urban neighborhoods for an extended period of time during these Millennials' young adulthood might influence their travel behavior in the long term. For instance, a recent survey in California has shown that Millennials' attitudes toward lifestyles and travel are different from those of Gen Xers (Circella et al., 2017). Staying in neighborhoods with a sufficient supply of transit, good walking environment, and additional opportunities for carpooling during young adulthood might encourage Millennials to explore various ways of travel (Smart & Klein, 2017). In addition, the future travel patterns of urban and suburban Millennials will be shaped by city-specific contexts (Delbosc et al., 2019). Testing whether such generational differences persist when the Millennials growing older can help to better understand the mechanisms determining the automobile travel behaviors of America's Millennials.

Nevertheless, considering the possibilities that such generational differences may persist over later life stages, there will be values for transportation planners to test the effectiveness of pilot projects targeting the new Millennial suburban dwellers at the moment. Projects such as bike sharing, park and ride, commuter rail, and vanpool programs may be more effective for Millennials than for those in previous generations. Many people migrate to suburban neighborhoods for reasons other than their preferred modes of traveling. Scholars have argued that neighborhoods that fostering multi-modalism may be undersupplied (Levine, 2006)). Although the total distance of automobile travel has bounced back recently in the U.S., adopting these programs may still be popular and successful.

From a researcher's perspective, there are a few important questions that will be particularly beneficial in extending this study. For example, testing whether the attitudes on and preferences for automobile travel differ across generations is valuable. Also, as mentioned above, researchers should continue testing to determine if these generational differences will persist over time. Due to a lack of data, this study is unable to control for variables such as gasoline prices, costs of alternative modes, built-environment patterns at the workplace, and level of congestion. Thus, future studies should include these variables. In addition, the current study focuses on associations rather than causations for the density-automobile relationships across generations. Future studies should consider building causal relationship between built environment and automobility. The present study focuses on a specific sub-sample in nationwide travel surveys: Millennial and Gen Xers living in metropolitan areas. Researchers interested in the aggregate passenger travel trends should consider examining generational differences among other age groups (e.g. older adults) and in rural areas.

# Appendix. Tables for sample sizes

**Table A1 – Sample sizes for Tables 6**

| Residential density (persons per square mile) | 16-28 | 29-41 | 42-54 | 55-67 | 68- |
|---|---|---|---|---|---|
| *1995* | | | | | |
| 0 to 100 | 700 | 1,328 | 1,167 | 739 | 516 |
| 100 to 500 | 1,617 | 2,864 | 2,643 | 1,490 | 1,022 |
| 500 to 1,000 | 1,043 | 1,796 | 1,544 | 977 | 750 |
| 1,000 to 2,000 | 1,397 | 2,358 | 1,980 | 1,248 | 1,070 |
| 2,000 to 4,000 | 2,196 | 3,235 | 2,860 | 1,734 | 1,673 |
| 4,000 to 10,000 | 3,321 | 4,525 | 3,725 | 2,437 | 2,240 |
| 10,000 to 25,000 | 1,348 | 1,668 | 1,238 | 812 | 681 |
| 25,000 - | 848 | 1,082 | 748 | 403 | 357 |
| Total | 12,470 | 18,856 | 15,905 | 9,840 | 8,309 |
| *2009* | | | | | |
| 0 to 100 | 1,351 | 1,677 | 3,632 | 4,485 | 3,416 |
| 100 to 500 | 2,880 | 3,993 | 8,047 | 9,346 | 7,195 |
| 500 to 1,000 | 1,840 | 2,706 | 4,694 | 5,615 | 4,965 |
| 1,000 to 2,000 | 2,846 | 4,069 | 7,619 | 8,642 | 8,341 |
| 2,000 to 4,000 | 4,255 | 5,935 | 11,141 | 12,637 | 12,687 |
| 4,000 to 10,000 | 5,486 | 7,128 | 12,777 | 13,646 | 13,692 |
| 10,000 to 25,000 | 1,560 | 1,972 | 2,871 | 2,850 | 2,630 |
| 25,000 - | 474 | 617 | 827 | 923 | 856 |
| Total | 20,692 | 28,097 | 51,608 | 58,144 | 53,782 |

**Table A2 – Sample sizes for Tables 9**

| Residential density (persons per square mile) | 24-36 | 37-49 | 50-62 | 63- |
|---|---|---|---|---|
| *2001* | | | | |
| 0 to 100 | 1,455 | 2,536 | 2,137 | 1,709 |
| 100 to 500 | 2,573 | 4,483 | 3,431 | 2,789 |
| 500 to 1,000 | 1,616 | 2,321 | 1,921 | 1,694 |
| 1,000 to 2,000 | 2,406 | 3,433 | 2,833 | 2,771 |
| 2,000 to 4,000 | 3,343 | 4,900 | 4,068 | 4,399 |
| 4,000 to 10,000 | 4,960 | 6,073 | 4,920 | 5,435 |
| 10,000 to 25,000 | 1,715 | 1,822 | 1,278 | 1,523 |
| 25,000 - | 1,100 | 1,027 | 801 | 779 |
| Total | 19,168 | 26,595 | 21,389 | 21,099 |
| *2017* | | | | |
| 0 to 100 | 1,685 | 2,171 | 4,369 | 6,457 |
| 100 to 500 | 3,486 | 4,591 | 7,926 | 10,996 |
| 500 to 1,000 | 2,659 | 3,133 | 4,717 | 6,856 |
| 1,000 to 2,000 | 4,059 | 4,561 | 6,984 | 9,872 |
| 2,000 to 4,000 | 6,340 | 6,904 | 10,301 | 13,821 |
| 4,000 to 10,000 | 9,052 | 8,642 | 12,066 | 14,875 |
| 10,000 to 25,000 | 3,066 | 2,302 | 2,693 | 3,182 |
| 25,000 - | 1,101 | 744 | 748 | 898 |
| Total | 31,448 | 33,048 | 49,804 | 66,957 |

**Funding**

This research did not receive any specific grant from funding agencies in the public, commercial, or not-for-profit sectors.

**Acknowledgements**

I would like to thank the valuable advice and feedback from Prof. Marlon G. Boarnet, Prof. Genevieve Giuliano, Prof. Dowell Myers, Prof. Julie Zissimopoulos, the Editor, and the three anonymous referees. I would also like to thank the Federal Highway Administration to provide the state of residence for the respondents in the 2001 National Household Travel Survey.

**Supplemental Material**

*Has the Relationship between Urban and Suburban Automobile Travel Changed across Generations? Comparing Millennials and Gen Xers in the United States*

Xize Wang

This document provides supporting information on the following topics:

## I. Analyses on walking, biking and transit trips

This section examines the number of daily biking, walking and transit trips for Millennials and Gen Xers. I group the three types of trips together in the analyses because of the relatively small magnitude of the average trip numbers (for instance, the average biking trip for the 16- to 28-year-old Gen Xers at the lowest residential density category is 0.002).

Generational differences exist for these "non-auto" trips across all the residential density categories. The two-sample t-tests in Table S1 show that the 16- to 28-year-old Millennials have higher frequencies of these "non-auto" trips than their Gen X counterparts. The outputs of Table S1 are also discussed in Section 4.1 of the main text.

In addition, the density-trip frequency associations are flatter for Millennial young adults. In other words, the differences in biking, walking and transit trips between urban and suburban Millennials are smaller than those of the Gen Xers. However, the magnitude of the variation of these trips are not comparable with those of automobile trips. The regression output for the 16-28-year-old sample, shown in Table S2, is also discussed in Section 4.2 of the main text.

I have repeated these analyses for the 24- to 36-year-old sample using the 2001, 2009 and 2017 NHTS data; the results are shown in Tables S3 and S4, respectively. In general, the generational difference patterns still hold. However, please note that for the regression model (Table S2), the generational difference in the magnitude of the association has decreased to 12% and is only significant at 10% level. The outputs of Tables S3 and S4 are also discussed in Section 5 of the main text.

**Table S1 – Difference in the travel patterns for the 16- to 28-year-old age group, by block-group level residential density**

| Residential density (persons per square mile) | 1995 (Gen Xers) | 2009 (Millennials) | Change | Change in % | p-value |
|---|---|---|---|---|---|
| *Average daily walking/biking/transit trips* | | | | | |
| 0 to 100 | 0.12 | 0.31 | 0.20 | 168.0% | 0.066 |
| 100 to 500 | 0.20 | 0.31 | 0.11 | 53.2% | 0.047 |
| 500 to 1,000 | 0.29 | 0.41 | 0.12 | 41.1% | 0.09 |
| 1,000 to 2,000 | 0.30 | 0.50 | 0.20 | 67.6% | 0.004 |
| 2,000 to 4,000 | 0.29 | 0.42 | 0.13 | 44.6% | 0.012 |
| 4,000 to 10,000 | 0.45 | 0.59 | 0.14 | 30.5% | 0.004 |
| 10,000 to 25,000 | 0.81 | 1.02 | 0.21 | 26.5% | 0.047 |
| 25,000 - | 2.10 | 1.61 | -0.49 | -23.4% | 0.013 |
| All | 0.50 | 0.59 | 0.09 | 17.0% | 0.002 |
| *Sample sizes* | | | | | |
| 0 to 100 | 700 | 1,351 | | N/A | |
| 100 to 500 | 1,617 | 2,880 | | | |
| 500 to 1,000 | 1,043 | 1,840 | | | |
| 1,000 to 2,000 | 1,397 | 2,846 | | | |
| 2,000 to 4,000 | 2,196 | 4,255 | | | |
| 4,000 to 10,000 | 3,321 | 5,486 | | | |
| 10,000 to 25,000 | 1,348 | 1,560 | | | |
| 25,000 - | 848 | 474 | | | |
| All | 12,470 | 20,692 | | | |

Note: Adjusted for personal weights. The p-values reported are from two-sample t-tests. Respondents not residing in MSAs or with daily personal VMT larger than 214 were excluded.

**Table S2 – Regression model for daily trips for the 16- to 28-year-old age group in 1995, 2001 and 2009**

| 16-28 years old sample | Negative binomial model for daily walking/biking/transit trips | |
|---|---|---|
| | Coef. | Std. Err. |
| Population density (1,000 people per square mile) | 0.054*** | (0.003) |
| Density-year interactions | | |
| Population density X 2001 | -0.006 | (0.004) |
| Population density X 2009 | -0.024*** | (0.004) |
| Survey year | | |
|   1995 | ref. | |
|   2001 | 0.345*** | (0.052) |
|   2009 | 0.613*** | (0.051) |
| Household income (in 1,000 2009 US Dollars) | -0.002*** | (0.000) |
| Driver status | -0.495*** | (0.034) |
| Household vehicles per person | -0.736*** | (0.038) |
| Age | -0.211*** | (0.054) |
| Age squared | 0.004*** | (0.001) |
| Female | -0.219*** | (0.041) |
| Having children | -0.442*** | (0.062) |
| Female X having children | 0.327*** | (0.073) |
| Worker | -0.228*** | (0.039) |
| Female X worker | 0.059 | (0.051) |
| Education | | |
|    Less than high school | ref. | |
|    High school/Associate degree | -0.067 | (0.043) |
|    Bachelor's degree | 0.354*** | (0.059) |
|    Graduate degree | 0.566*** | (0.084) |
|    Not available | 0.016 | (0.159) |
| Race of household head | | |
|    Non-Hispanic white | ref. | |
|    Non-Hispanic black | 0.165*** | (0.048) |
|    Non-Hispanic Asian/Pacific Islander | -0.031 | (0.062) |
|    Hispanic | -0.146*** | (0.043) |
|    Other races | 0.106 | (0.077) |
| Population of metropolitan area | | |
|    Lower than 250,000 | ref. | |
|    250,000 - 499,999 | 0.120** | (0.050) |
|    500,000 - 999,999 | 0.097* | (0.057) |
|    1,000,000 - 2,999,999 | 0.140*** | (0.048) |
|    3 million or higher | 0.120** | (0.048) |
| State fixed effects | | |
|    Alabama | ref. | |
|    Alaska | 1.210*** | (0.427) |
|    … | | |
|    Wyoming | 1.441** | (0.664) |
| Constant | 1.227* | (0.627) |
| N | 44,010 | |
| Pseudo R2 | 0.050 | |

Note: Standard errors in parentheses, *, **, *** indicate p<0.1, p<0.05 and p<0.01, respectively. Respondents not residing in MSAs or with personal VMT larger than 214 were excluded. The 16- to 28-year-olds in 2009 belong to the Millennial generation, and the 16- to 28-year-olds in 1995 belong to Generation X.

**Table S3 – Differences in the travel patterns for the 24- to 36-year-old age group, by block-group level residential density**

| Residential density (persons per square mile) | 2001 (Gen Xers) | 2017 (Millennials) | Change | Change in % | p-value |
|---|---|---|---|---|---|
| *Average daily walking/biking/transit trips* | | | | | |
| 0 to 100 | 0.22 | 0.29 | 0.07 | 32.3% | 0.382 |
| 100 to 500 | 0.25 | 0.22 | -0.04 | -14.4% | 0.404 |
| 500 to 1,000 | 0.24 | 0.38 | 0.13 | 55.6% | 0.044 |
| 1,000 to 2,000 | 0.27 | 0.38 | 0.11 | 41.5% | 0.021 |
| 2,000 to 4,000 | 0.32 | 0.51 | 0.19 | 59.2% | <0.001 |
| 4,000 to 10,000 | 0.46 | 0.58 | 0.12 | 26.9% | 0.003 |
| 10,000 to 25,000 | 0.78 | 1.22 | 0.44 | 56.5% | <0.001 |
| 25,000 - | 2.06 | 2.58 | 0.52 | 25.2% | 0.001 |
| All | 0.52 | 0.73 | 0.21 | 39.6% | <0.001 |
| *Sample sizes* | | | | | |
| 0 to 100 | 1,455 | 1,685 | | | |
| 100 to 500 | 2,573 | 3,486 | | | |
| 500 to 1,000 | 1,616 | 2,659 | | | |
| 1,000 to 2,000 | 2,406 | 4,059 | | N/A | |
| 2,000 to 4,000 | 3,343 | 6,340 | | | |
| 4,000 to 10,000 | 4,960 | 9,052 | | | |
| 10,000 to 25,000 | 1,715 | 3,066 | | | |
| 25,000 - | 1,100 | 1,101 | | | |
| All | 19,168 | 31,448 | | | |

Note: Adjusted for personal weights. The p-values reported are from two-sample t-tests. Respondents not residing in MSAs or with daily personal VMT larger than 214 were excluded.

**Table S4 – Regression model for daily trips for the 24- to 36-year-old age group in 2001, 2009 and 2017**

| 24-36 years old sample | Negative binomial model for daily walking/biking/transit trips | |
|---|---|---|
| | Coef. | Std. Err. |
| Population density (1,000 people per square mile) | 0.052*** | (0.003) |
| Density-year interactions | | |
| Population density X 2009 | -0.007 | (0.004) |
| Population density X 2017 | -0.006* | (0.003) |
| Survey year | | |
| 2001 | ref. | |
| 2009 | 0.216*** | (0.043) |
| 2017 | 0.249*** | (0.039) |
| Household income (in 1,000 2009 US Dollars) | -0.001*** | (<0.001) |
| Driver status | -0.695*** | (0.043) |
| Household vehicles per person | -0.511*** | (0.030) |
| Age | 0.165*** | (0.052) |
| Age squared | -0.003*** | (0.001) |
| Female | -0.066 | (0.056) |
| Having children | -0.503*** | (0.036) |
| Female X having children | 0.201*** | (0.043) |
| Worker | -0.142*** | (0.049) |
| Female X worker | -0.019 | (0.058) |
| Education | | |
|   Less than high school | ref. | |
|   High school/Associate degree | -0.072 | (0.057) |
|   Bachelor's degree | 0.469*** | (0.060) |
|   Graduate degree | 0.738*** | (0.063) |
|   Not available | 0.302* | (0.179) |
| Race of household head | | |
|   Non-Hispanic white | ref. | |
|   Non-Hispanic black | -0.073 | (0.045) |
|   Non-Hispanic Asian/Pacific Islander | -0.292*** | (0.044) |
|   Hispanic | -0.199*** | (0.036) |
|   Other races | -0.099 | (0.061) |
| Population of metropolitan area | | |
|   Lower than 250,000 | ref. | |
|   250,000 - 499,999 | 0.050 | (0.041) |
|   500,000 - 999,999 | 0.040 | (0.043) |
|   1,000,000 - 2,999,999 | 0.096** | (0.039) |
|   3 million or higher | 0.078** | (0.038) |
| State fixed effects | | |
|   Alabama | ref. | |
|   Alaska | 1.066*** | (0.354) |
|   … | | |
|   Wyoming | 1.076** | (0.467) |
| Constant | -3.554*** | (0.812) |
| N | 68,900 | |
| Pseudo R-squared | 0.0401 | |

Note: Standard errors in parentheses, *, **, *** indicate p<0.1, p<0.05 and p<0.01, respectively. Respondents not residing in MSAs or with personal VMT larger than 214 were excluded. The 24- to 36-year-olds in 2017 belong to Millennials, and the 24- to 36-year-olds in 2001 belong to Generation X.

## II. Density-automobility relationships across generations, interacted by gender

This section examines whether the generational difference of the density-automobility relationships varies by gender. The results of the regression models with density-automobility-gender interaction terms are shown in Table S5 (the 16- to 28-year-old sample) and Table S6 (the 24- to 36-year-old sample) below.

In general, the generational differences of the density-automobility relationships does not vary by gender. The only exception is the daily personal VMT for the 24- to 36-year-old sample, as shown in Table S6. This table shows that the density-personal VMT associations are flatter for female Millennials than male Millennials for the 24- to 36-year-old age group.

The results of these gender-interaction models for the 16-to 28-year-old sample are discussed in Section 4.2 of the main text. The results of these gender-interaction models for the 24- to 36-year-old sample are discussed in Section 5 of the main text.

Table S5 – Regression with interaction terms with gender: the 16- to 28-year-old sample

| 16-28-year-old sample | Tobit model for daily personal VMT | | Negative binomial model for daily car trips | |
|---|---|---|---|---|
| | Coef. | Std. Err. | Coef. | Std. Err. |
| Population density (1,000 people per square mile) | -1.043*** | (0.057) | -0.015*** | (0.001) |
| Density-year interactions | | | | |
|    Population density X 2001 | 0.044 | (0.084) | 0.002 | (0.002) |
|    Population density X 2009 | 0.318*** | (0.088) | 0.008*** | (0.002) |
| Density-year-gender interactions | | | | |
|    Population density X 2001 X female | -0.022 | (0.080) | -0.002 | (0.002) |
|    Population density X 2009 X female | 0.005 | (0.088) | -0.001 | (0.002) |
| Survey year | | | | |
|    1995 | ref. | | ref. | |
|    2001 | -3.030*** | (0.765) | -0.172*** | (0.016) |
|    2009 | -8.902*** | (0.748) | -0.441*** | (0.016) |
| Household income (in 1,000 2009 US Dollars) | 0.047*** | (0.005) | 0.001*** | (<0.001) |
| Driver status | 16.862*** | (0.608) | 0.665*** | (0.014) |
| Household vehicles per person | 10.616*** | (0.572) | 0.193*** | (0.012) |
| Age | -0.081 | (0.834) | -0.129*** | (0.018) |
| Age squared | 0.003 | (0.018) | 0.003*** | (0.000) |
| Female | 1.931*** | (0.733) | 0.092*** | (0.016) |
| Having children | 9.044*** | (0.902) | 0.140*** | (0.019) |
| Female X having children | -6.725*** | (1.064) | 0.032 | (0.022) |
| Worker | 9.739*** | (0.635) | 0.219*** | (0.014) |
| Female X worker | -1.546* | (0.816) | -0.008 | (0.018) |
| Education | | | | |
|   Less than high school | ref. | | ref. | |
|   High school/Associate degree | 3.615*** | (0.669) | 0.033** | (0.014) |
|   Bachelor's degree | 6.619*** | (0.899) | 0.058*** | (0.019) |
|   Graduate degree | 5.864*** | (1.313) | 0.052* | (0.028) |
|   Not available | 2.984 | (2.524) | 0.110** | (0.055) |
| Race of household head | | | | |
|   Non-Hispanic white | ref. | | ref. | |
|   Non-Hispanic black | -3.912*** | (0.807) | -0.092*** | (0.017) |
|   Non-Hispanic Asian/Pacific Islander | -1.046 | (1.019) | -0.067*** | (0.022) |
|   Hispanic | -1.260* | (0.677) | -0.022 | (0.015) |
|   Other races | -1.497 | (1.233) | -0.046* | (0.026) |
| Population of metropolitan area | | | | |
|   Lower than 250,000 | ref. | | ref. | |
|   250,000 - 499,999 | 1.484** | (0.734) | -0.036** | (0.015) |
|   500,000 - 999,999 | 1.530* | (0.831) | -0.039** | (0.017) |
|   1,000,000 - 2,999,999 | 0.914 | (0.716) | -0.045*** | (0.015) |
|   3 million or higher | 0.993 | (0.712) | -0.107*** | (0.015) |
| State fixed effects | | | | |
|   Alabama | ref. | | ref. | |
|   Alaska | 2.325 | (6.297) | 0.181 | (0.131) |
|   … | | | | |
|   Wyoming | -9.429 | (10.045) | 0.338* | (0.197) |
| Constant | -0.142 | (9.399) | 1.987*** | (0.200) |
| N | 44,010 | | 44,010 | |
| Pseudo R-squared | 0.0171 | | 0.0398 | |

Note: Standard errors in parentheses, *, **, *** indicate p<0.1, p<0.05 and p<0.01, respectively. Respondents not residing in MSAs or with personal VMT larger than 214 were excluded. The 16- to 28-year-olds in 2009 belong to the Millennial generation, and the 16- to 28-year-olds in 1995 belong to Generation X.

Table S6 – Regression with interaction terms with gender: the 24- to 36-year-old sample

| 24-36-year-old sample | Tobit model for daily personal VMT | | Negative binomial model for daily car trips | |
|---|---|---|---|---|
| | Coef. | Std. Err. | Coef. | Std. Err. |
| Population density (1,000 people per square mile) | -1.072*** | (0.045) | -0.013*** | (0.001) |
| Density-year interactions | | | | |
|   Population density X 2009 | 0.142* | (0.081) | 0.002 | (0.002) |
|   Population density X 2017 | 0.118* | (0.067) | 0.003** | (0.001) |
| Density-year-gender interactions | | | | |
|   Population density X 2009 X female | -0.026 | (0.085) | 0.002 | (0.002) |
|   Population density X 2017 X female | 0.182*** | (0.064) | 0.002 | (0.001) |
| Survey year | | | | |
|   2001 | ref. | | ref. | |
|   2009 | -4.731*** | (0.607) | -0.166*** | (0.012) |
|   2017 | -7.797*** | (0.544) | -0.217*** | (0.011) |
| Household income (in 1,000 2009 US Dollars) | 0.032*** | (0.004) | <0.001 | (<0.001) |
| Driver status | 26.339*** | (0.791) | 0.908*** | (0.018) |
| Household vehicles per person | 10.329*** | (0.442) | 0.131*** | (0.009) |
| Age | -0.779 | (0.764) | -0.018 | (0.015) |
| Age squared | 0.011 | (0.013) | <0.001 | (<0.001) |
| Female | 3.246*** | (0.931) | 0.119*** | (0.019) |
| Having children | 8.986*** | (0.533) | 0.160*** | (0.010) |
| Female X having children | -4.151*** | (0.643) | 0.066*** | (0.013) |
| Worker | 14.834*** | (0.777) | 0.201*** | (0.016) |
| Female X worker | -3.342*** | (0.912) | -0.047** | (0.019) |
| Education | | | | |
|   Less than high school | ref. | | ref. | |
|   High school/Associate degree | 2.866*** | (0.868) | 0.100*** | (0.018) |
|   Bachelor's degree | 5.128*** | (0.910) | 0.158*** | (0.018) |
|   Graduate degree | 3.377*** | (0.956) | 0.149*** | (0.019) |
|   Not available | -1.180 | (2.914) | -0.065 | (0.061) |
| Race of household head | | | | |
|   Non-Hispanic white | ref. | | ref. | |
|   Non-Hispanic black | -0.002 | (0.673) | -0.006 | (0.013) |
|   Non-Hispanic Asian/Pacific Islander | -2.907*** | (0.681) | -0.108*** | (0.014) |
|   Hispanic | -0.210 | (0.524) | -0.006 | (0.010) |
|   Other races | 0.123 | (0.907) | -0.014 | (0.018) |
| Population of metropolitan area | | | | |
|   Lower than 250,000 | ref. | | ref. | |
|   250,000 - 499,999 | -0.057 | (0.585) | -0.036*** | (0.011) |
| 500,000 - 999,999 | 1.697*** | (0.625) | -0.016 | (0.012) |
| 1,000,000 - 2,999,999 | 2.312*** | (0.564) | -0.016 | (0.011) |
| 3 million or higher | 3.613*** | (0.557) | -0.081*** | (0.011) |
| State fixed effects | | | | |
|   Alabama | ref. | | ref. | |
|   Alaska | 1.246 | (4.775) | -0.003 | (0.093) |
|   … | | | | |
|   Wyoming | -16.868** | (6.555) | -0.176 | (0.129) |
| Constant | -5.095 | (11.802) | 0.208 | (0.231) |
| N | 68,900 | | 68,900 | |
| Pseudo R-squared | 0.0131 | | 0.0279 | |

Note: Standard errors in parentheses, *, **, *** indicate p<0.1, p<0.05 and p<0.01, respectively. Respondents not residing in MSAs or with personal VMT larger than 214 were excluded. The 24- to 36-year-olds in 2017 belong to Millennials, and the 24- to 36-year-olds in 2001 belong to Generation X.